\documentclass[fdp,fleqn]{w-art}
\usepackage{times}
\usepackage{w-thm}
\usepackage{amssymb}
\usepackage[]{graphicx}

\usepackage{comment}

\def\bra{\langle}
\def\ket{\rangle}
\def\beq{\begin{equation}}
\def\eeq{\end{equation}}
\newcommand{\C}[1]{\mathcal{#1}}
\def\ov{\overline}

\begin{document}
%%    The information for the title page will be placed between
%%    \begin{document} and \maketitle. The order of most entries
%%    is determined by the class file and can not be changed by
%%    rearranging them. The maketitle command follows after the
%%    abstract.
%%
%%    Most of the following commands will be completed by the publisher.
%%
%%    The copyrightyear is defined in the .clo file as the first argument
%%    of the copyrightinfo command. If the copyrightyear differs from that
%%    value it might be adjusted by the following definition:
%%
%% \renewcommand{\copyrightyear}{2003}% uncomment to change the copyrightyear.
%%
%\DOIsuffix{theDOIsuffix}
%%
%% issueinfo for header and copyright line
%\Volume{55}
%\Issue{1}
%\Month{01}
%\Year{2007}
%%
%%    First and last pagenumber of the article. If the option
%%    'autolastpage' is set (default) the second argument may be left empty.
\pagespan{1}{}
%%
%%    Dates will be filled in by the publisher. The 'reviseddate' and
%%    'dateposted' (Published online) entry may be left empty.
%\Receiveddate{9 February 2010}
%\Reviseddate{31 January 2010}
%\Accepteddate{1 February 2010}
%\Dateposted{2 February 2010}
%%
\keywords{Heterotic strings, type II strings, string phenomenology.}
\subjclass[pacs]{11.25.Mj, 11.25.Wx}

%% \pretitle{Editor's Choice}

%% We have a short and a long form for the title. The short form
%% (optional argument) goes into the running head.

\title[Light hidden-sector U(1)s]{Light hidden-sector U(1)s in string compactifications}

%% Please do not enter footnotes or \inst{}-notes into the optional
%% argument of the author command. The optional argument will go into
%% the header.  If there is only one address the marker \inst{x} may be
%% omitted.

%% Information for the first author.
\author[M. Goodsell]{Mark Goodsell\inst{1,}%
  \footnote{E-mail:~\textsf{mark.goodsell@desy.de}}}
\address[\inst{1}]{Deutsches Elektronen-Synchrotron DESY, Notkestrasse 85, D-22607 Hamburg, Germany}
%%
%%    Information for the second author
\author[A. Ringwald]{Andreas Ringwald\inst{1,}%
  \footnote{Corresponding author\quad E-mail:~\textsf{andreas.ringwald@desy.de}}}
%%
%%    \dedicatory{This is a dedicatory.}
\begin{abstract}
We review the case for light U(1) gauge bosons in the hidden-sector of
heterotic and type II string compactifications, present
estimates of the size of their kinetic mixing with the visible-sector hypercharge U(1),
and discuss their possibly very interesting phenomenological consequences in particle physics
and cosmology.
\end{abstract}
%% maketitle must follow the abstract.
\maketitle                   % Produces the title.

%% If there is not enough space inside the running head
%% for all authors including the title you may provide
%% the leftmark in one of the following three forms:

%% \renewcommand{\leftmark}
%% {F. Author: A short title}

%% \renewcommand{\leftmark}
%% {F. Author and S. Author: A short title}

%% \renewcommand{\leftmark}
%% {F. Author et al.: A short title}

%% \tableofcontents  % Produces the table of contents.
\section{Introduction}
There are presently three major paradigms for string phenomenology: compactifications of the
heterotic string, of type II strings with D-branes, and of $F$-theory.
And in all these scenarios it appears that in attempts to obtain the
desired (minimal supersymmetric) standard model
almost inevitably there arises also a hidden sector of gauge bosons and matter particles
which interacts with the visible standard model sector only very weakly
because there are no light messenger states charged under both gauge sectors.

For a long time, it was assumed that direct effects associated with the hidden sector
are unobservably small at our available energies, since the interactions
between standard model and hidden sector particles occur via
operators of mass dimension bigger than four that arise from integrating out the messenger fields.
Correspondingly, one expected a huge suppression of the effective couplings by inverse powers
of the mass scale of the latter.
However, these considerations were neglecting the possible existence of
light hidden Abelian gauge bosons. Contrary to non-Abelian gauge bosons, hidden U(1) gauge bosons
can mix kinetically with the visible sector hypercharge U(1) gauge boson, leading to a
term of mass dimension four,
$\sim \chi\, d A \wedge \star d V$, in the low-energy effective Lagrangian,
where $dA$ ($d V$) is the hypercharge (hidden) U(1) field strength tensor.
Correspondingly, the dimensionless kinetic mixing parameter
$\chi$ is not necessarily parametrically small, even if the messengers are ultra-heavy.
Moreover, on the mass dimension two level, there is also the possibility of
St\"uckelberg mass mixing between hidden U(1)s and the hypercharge U(1).

Various phenomenologically very interesting effects can arise from kinetic mixing,
opening a window on the hidden sector.
The least effect occurs if the hidden photon is massless and there is no light hidden matter.
In this case, the kinetic mixing can be removed by a rotation in field space, its only effect
then being a multiplicative renormalization of
the hypercharge gauge coupling, shifting it away from its high-energy
value~\cite{Redondo:2008zf}. However, in supersymmetry, the hidden photon is accompanied by a hidden
photino, which can mix with the MSSM neutralinos and may thus lead to more directly observable effects,
even if the hidden photon stays massless, ranging from hidden photino 
superWIMP ($\chi\sim 10^{-11}\div 10^{-8}$) or decaying ($\chi\sim 10^{-23}$) dark matter to
direct hidden photino production at the LHC~\cite{Ibarra:2008kn,Ibarra:2009bm,Arvanitaki:2009hb}.
Important effects could also arise if there were
light hidden-sector matter charged under the hidden U(1), since the latter has
a mini-charge $\sim \chi g_h/e$, where $g_h$ is the hidden U(1) gauge coupling, under the
visible U(1)~\cite{Holdom:1985ag}, the latter being constrained to be $\lesssim 10^{-14}$ by stellar energy loss
considerations~\cite{Davidson:2000hf}.
If the hidden photon has a mass,
the hidden U(1) forces can be attacked on even more fronts.
The best sensitivity for kinetic mixing of hidden U(1)s arises in this case
in the search for signatures of photon $\leftrightarrow$
hidden photon oscillations in the CMB~\cite{Jaeckel:2008fi,Mirizzi:2009iz}, in laser photon regeneration
experiments~\cite{Ahlers:2007rd}
(both together excluding $\chi\gtrsim 10^{-7}$ for masses between
$10^{-14}\,{\rm eV}$ and $10\,{\rm meV}$), in the sun, in stars, and in cosmology~\cite{Redondo:2008aa,Redondo:2008ec}
(excluding values down to $\chi\gtrsim 10^{-14}$ in the
$10\,{\rm meV}$ to $1\,{\rm MeV}$ mass range). For higher masses, the hidden photon
may be produced and searched for in fixed-target (sensitivity down to $\chi\sim10^{-7}$)
or collider experiments~\cite{Essig:2009nc,Bjorken:2009mm}
(sensitivity down to $\chi\sim 10^{-3}$). Intriguingly,
a hidden photon in the mass range from $10\,{\rm MeV}$ to $1\,{\rm GeV}$, mixing
with a strength $\chi$ between $10^{-7}$ and $10^{-3}$, may provide a unified description
of the anomalous positron excesses observed in recent cosmic ray data by
PAMELA
(in terms of hidden sector dark matter annihilation
into pairs of hidden photons)
and the annual modulation signal observed by the
dark matter search experiment DAMA,
without running into contradictions with the null results of other direct dark matter
search experiments~\cite{ArkaniHamed:2008qn}. For masses above $\sim 100$~GeV, the
hidden photon acts as a $Z^\prime$ candidate
 and may be searched for in Drell-Yan processes
at the LHC, with a sensitivity in the $\chi\sim 10^{-2}$ range~\cite{Kumar:2006gm,Feldman:2007wj}.

Hidden U(1) gauge factors are ubiquitous in string compactifications: in the
heterotic string, they can be found e.g. in the standard embedding
by breaking the hidden E$_8$. In type II/$F$ theories, there are so-called RR U(1)s, arising
as zero modes of closed string RR form fields, which have no matter charged under them.
Finally, realizing U(1) gauge bosons as
excitations of space-time filling D-branes wrapping cycles in the extra
dimensions, extra U(1)s can be hidden from the hypercharge U(1), if the corresponding
D-branes are separated in the compact space.
We shall discuss kinetic mixing and masses in all of these cases.

\section{Heterotic Hidden U(1)s}

In the heterotic string every group factor arises from the breaking of E$_8 \times $E$_8$ or Spin(32)/Z$_2$. There are two complimentary methods of determining the kinetic mixing: one is direct string loop calculation and the other is dimensional reduction of the Green-Schwarz counter-term. The first approach is only possible when there is a CFT description available, such as for toroidal orbifolds, where once the condition of no light messengers
was imposed the mixing in specific examples was found to be zero~\cite{Dienes:1996zr}.
The second approach is particular to the use of vacuum expectation values for the hidden U(1) fields,
where non-zero mixing was found~\cite{Lukas:1999nh,Blumenhagen:2005ga}.

In unwarped heterotic compactifications, all the gauge couplings are related to the dilaton  and the
only mass scales are set by $M_s\sim g M_{P}$. Thus, generically, one expects
$\chi\sim g_Y g_h/(16\pi )^2\sim g_Y^2/(16\pi^2 )\sim {\rm few}\times 10^{-4}$ for the mixing between the hypercharge
and a hidden U(1) generated in the supersymmetric limit.
Typically, in those models supersymmetry breaking is mediated by gravity,
which leads to a hidden photino mass of the same order as those of the MSSM neutralinos. Overproduction in the early
universe then may exclude such large mixing~\cite{Ibarra:2008kn}, although this could be avoided if, for
example, there is hidden matter into which the hidden photino can decay~\cite{Arvanitaki:2009hb}.
Alternatively, kinetic mixing can instead be
generated only once supersymmetry is broken. The lowest-order operators that contribute are schematically
\begin{align}
\Delta \mathcal{L} \supset  \int d^4 \theta\ W^a W^b \left[ \frac{S + \ov{S}}{M_s^2}
+\frac{{\mathcal{D}}^2 (\ov{S} + S)^2}{M_s^4}  \right] + {\rm c.c.} ,
\label{secondorder}
\end{align}
where $W_a, W_b$ are the field strength superfields for the two U(1) gauge fields, $S$ is the dilaton and
$\mathcal{D}$ a superspace derivative.
For the $F$-term expectation value of the dilaton $S$ at $10^{21}$\,GeV$^2$, as typical for gravity mediation,
these operators give mixings of order $(g_Y g_h)/(16\pi^2)\cdot (F_s/M_s^2)\sim 10^{-17}$ and
$(g_Y g_h)/(16\pi^2)\cdot (F_s/M_s^2)^2\sim 10^{-30}$, respectively, since $M_s\sim 10^{17}$\,GeV in the
weakly coupled heterotic string.
However, a calculation in field theory leads us to expect the first operator to vanish.
This suggests to obtain phenomologically interesting  values in this case, we should allow for intermediate mass
messengers.

\section{RR U(1)s}

In type IIB string theory, the dimensional reduction of the Ramond-Ramond $4$-form $C_4$ yields $h^{3}$ four-dimensional
vectors\footnote{There is a similar story for $C_3/C_5$ in type IIA, but this is analagous and for simplicity we shall focus on type IIB.}. These U(1)s do not possess any charged matter and in Calabi-Yau compactifications have no axionic couplings, and are thus hidden and massless. They can become massive in \emph{non-K\"ahler} compactifications \cite{Grimm:2008ed} by acquiring axionic couplings; in this case they may also exhibit \emph{mass mixing} with the hypercharge. However, if we restrict our attention to the Calabi-Yau case, the only way that they can be detected is by their kinetic mixing with the hypercharge, supported on branes.

The kinetic mixing of RR U(1)s with D7-branes was calculated in \cite{Jockers:2004yj}, and with D5-branes in \cite{Grimm:2008dq}. In the first case, it is necessary for there to be Wilson line moduli to obtain non-zero mixing, while in the second deformation moduli of the D5-branes are required. This means that the branes are not rigid prior to moduli stabilisation; the brane moduli appear as massles adjoint fields. Supposing the hypercharge U(1) being localized on a wrapped D7 brane,
the kinetic mixing with a RR U(1) arises from the Chern-Simons coupling and can be inferred from the low energy effective
action which reads, after dimensional reduction~\cite{Jockers:2004yj},
\begin{align}
S_{\rm YM} \supset& -\int_4 2\pi \frac{1}{2}\mathrm{Im} (\C{M})_{\alpha \beta} d V^\alpha \wedge \star_4 d V^\beta - \frac{1}{\pi} C_{\alpha \beta} \bigg( (a^\beta + \ov{a}^\beta) dV^\alpha \wedge \star_4 \, dA \bigg),
\end{align}
where  $(\C{M})_{\alpha \beta}, C_{\alpha \beta}$ are $\C{O}(1)$ dimensionless matrices independent of the overall volume. The quantities $a^\beta, \ov{a}^\beta$ are numbers parametrising the magnitude of the Wilson line moduli (which must therefore be stablised at non-zero values to obtain non-zero mixing) which may in principle have some very weak volume dependence via their kinetic term, which scales as $\sim g_Y^{-1}$.
Therefore the magnitude of the physical kinetic mixing between these RR U(1)s and the hypercharge depends on how the Wilson line moduli are stabilised at a non-zero value. If it is via a stringy mechanism, the mixing is expected to be quite large,
$\chi \sim  (g_Y)^{3/2}/(\pi\sqrt{2})\sim 10^{-3}$. On the other hand, if it occurs via a field theory mechanism (for example if $a$ is the singlet of the NMSSM, $N$) then much smaller values would be obtained, $\chi \sim  g_Y/(\pi\sqrt{2})
\cdot \bra N \ket/M_s\sim 10^{-15}\, (10^{15}\ \mathrm{GeV}/M_s)$. Nevertheless, it seems that effects of RR U(1)s
will be hard to detect, because the hidden photinos  are very light:
since the gauge coupling does not depend upon the K\"ahler moduli, they typically do not obtain a mass at the tree level, only acquiring one through their mixing. If the adjoints $a$ acquire an $F$-term vev, then mass mixing terms can arise - otherwise the photinos can only obtain a mass through gravity loops.

\section{Hidden U(1)s on D-Branes}

The simplest way to engineer kinetic mixing in type II string theories is to consider brane-antibrane kinetic mixing on a toroidal background~\cite{Abel:2003ue,Abel:2006qt,Abel:2008ai}. Indeed, anti-D3 branes can be introduced in IIB compactifications in order to uplift to a deSitter vacuum. The mixing can then be calculated directly using CFT techniques or alternatively estimated using supergravity. Since antibranes violate supersymmetry at the string scale, one cannot appeal to holomorphy to constrain the mixing; the direct computation of \cite{Abel:2003ue} showed the mixing between a $p$-brane and an anti-D3 brane on a homogeneous six-torus of radius $R$ to be
$\chi_{ab} \sim g_a g_b(2\pi R)^{p-7}/(16\pi^2)  \sim g_s (2\pi R)^{(p-11)/2}/(16\pi^2)
\sim  g_s^{(p-3)/8} \C{V}^{(p-11)/12}/(16\pi^2)$,
where $\C{O}(1)$ terms depending only upon the complex structure have been dropped.
The case of most phenomenological interest is the mixing between a collapsed D7 brane, on which the
hypercharge U(1) is localized, and a hidden $\ov{{\rm D}3}$ brane. It exhibits a large volume suppression\footnote{Here $t_i$ are the volumes of two-cycles which support KK modes that the branes couple to, and $f \propto t_i$ as $t_i \rightarrow \infty$.},
$\chi \sim f(t_i)/(16\pi^2\C{V}) \sim 10^{-2}\, \C{V}^{-1} \div 10^{-2}\, \C{V}^{-2/3}$,
yielding $\sim 10^{-8} \div 10^{-6}$, for GUT scale strings (cf. $\C{V}\sim M_P^2/M_s^2$).
The St\"uckelberg masses of hidden U(1)s supported on $\ov{\rm D3}$ branes turn out to be zero, since the massless modes of the axions that generate the mixing (the two forms $B_2$ and $C_2$) are projected out of the spectrum by the orientifold (leaving only KK modes).
Therefore, these hidden U(1)s are massless. Moreover, similar to the RR U(1)s, their superpartners do not acquire
a tree level mass, rendering this uplifting part of the hidden sector quite invisible.

We now turn our attention to supersymmetric mixing.
Supersymmetry can be used to determine the moduli dependence of the kinetic mixing, since it must appear in the gauge kinetic terms as a holomorphic parameter~\cite{Benakli:2009mk,Goodsell:2009xc}. The closed string K\"ahler moduli $T_\alpha$ transform under Peccei-Quinn symmetries and they can therefore only enter as exponentials~\cite{Akerblom:2007uc}.
However, they also depend upon the inverse string coupling
and consequently an exponential
dependence would be non-perturbative. Thus they \emph{cannot} enter at one-loop. Generically, the complex structure moduli enter the holomorphic kinetic mixing in polynomial or exponential form and will typically be numbers of order 
one (for a T-dual example in a IIA model, see Ref.~\cite{Lust:2003ky}), although some may be exponentially small 
at the end of a warped throat~\cite{Abel:2008ai}. We thus conclude that generically the physical kinetic mixing is
$\chi_{ab} \sim g_a g_b/(16\pi^2)$~\cite{Goodsell:2009xc}.
Then for hidden U(1)s with similar gauge coupling as the hypercharge we will always obtain mixing $\chi \sim 10^{-4}$. However, in IIB compactifications there is also the possibility that the hidden gauge group wraps a large cycle, and so obtains a hyperweak gauge coupling $g_h \sim \C{V}^{-1/3}$~\cite{Burgess:2008ri}. In such cases we may then obtain much smaller values of the kinetic mixing, which correlate with the volume and thus the string scale, $\chi\sim (M_s/M_P)^{2/3}/(16 \pi^2)\sim
10^{-8}\div 10^{-4}$, for $M_s\sim 10^{10}\div 10^{16}$\,GeV~\cite{Goodsell:2009xc}.

A St\"uckelberg mass is induced in case that the gauge boson eats an axion, e.g.  $m_{\rm Stueck}\sim M_s^2/M_{P}$,
provided that the brane wraps a cycle even under the orientifold projection.
This will lead to a mass in the tens of GeV range, for an intermediate
string scale $M_s\sim 10^{10}$\,GeV, with a phenomenologically quite interesting mixing of order $\sim 10^{-8}$ (cf. above).
The predictions for masses arising from the Higgs mechanism are less precise; however, they can also be
tiny if the supersymmetry breaking scale in the hidden sector is much smaller than in the visible sector.

\section{Conclusions}

Compactifications of heterotic and type II strings allow for a variety of
hidden-sector U(1)s, kinetically mixing with the hypercharge U(1). These could play
a role in a number of phenomena, in particular connected to dark matter.
Near future laboratory experiments may probe the hidden sector of string theory!

\begin{acknowledgement}
MG acknowledges support from the German Science Foundation (DFG) under SFB 676.
\end{acknowledgement}

\end{document}